\documentclass[twocolumn,showpacs,preprintnumbers,amssymb,aps,floatfix,draft]{revtex4}
\usepackage{epsf}
\usepackage{times}
\newcommand{\trace}{{\mathrm Tr }}
\newcommand{\twovec}[2]{\left(\begin{array}{c} #1 \\ #2 \end{array}\right)}
\newcommand{\twomat}[4]{\left(\begin{array}{cc} #1 & #2 \\ #3 & #4\end{array}\right)}
\newcommand{\qed}{\hfill$\square$\par\vskip24pt}

\newcommand{\R}{{\mathbb{R}}}
\newcommand{\identity}{\openone}
\newcommand{\be}{$$}
\newcommand{\ee}{$$}
\newcommand{\bea}{\begin{eqnarray}}
\newcommand{\eea}{\end{eqnarray}}
\newcommand{\beas}{\begin{eqnarray*}}
\newcommand{\eeas}{\end{eqnarray*}}

\newtheorem{theorem}{Theorem}
\newtheorem{lemma}{Lemma}
\newtheorem{corollary}{Corollary}

\begin{document}

\title{Entanglement Properties of the Harmonic Chain}
\author{K.\ Audenaert}
\email{k.audenaert@ic.ac.uk}
\author{J.\ Eisert}
\email{j.eisert@ic.ac.uk}
\author{M.B.\ Plenio}
\email{m.plenio@ic.ac.uk}

\affiliation{QOLS, Blackett Laboratory, Imperial College of
Science, Technology and Medicine, London, SW7 2BW, UK}

\author{R.F.\ Werner}
\email{R.Werner@tu-bs.de}
\affiliation{Institut f\"ur Mathematische
Physik, TU Braunschweig, Mendelssohnstra{\ss}e 3, 38106
Braunschweig, Germany}

\date{\today}

\begin{abstract}
We study the entanglement properties of a closed chain of harmonic
oscillators that are coupled via a translationally invariant
Hamiltonian, where the coupling acts only on the position
operators. We consider the ground state and thermal states of this
system, which are Gaussian states. The entanglement properties of
these states can be completely characterized analytically when one
uses the logarithmic negativity as a measure of entanglement.
\end{abstract}


\maketitle
\section{Introduction} \label{sec_intro}

Quantum entanglement is possibly the most intriguing property of
states of composite quantum systems. It manifests itself in
correlations of measurement outcomes that are stronger than
attainable in any classical system. The renewed interest in a
general theory of entanglement in recent years is largely due to
the fact that entanglement is conceived as the key resource in
protocols for quantum information processing.
Initial investigations focused on the properties of
bipartite entanglement of finite dimensional systems
such as two-level systems.
In fact, significant progress has been made,
and our understanding of the entanglement of
such systems is quite well developed
\cite{Entanglement}.
A natural next step is the extension of these
investigations to multi-partite systems. Unfortunately, the study
of multi-partite entanglement suffers from a proliferation of
different types of entanglement already
in the pure state case \cite{MREGS},
and even less is known
about the mixed state case. For example, necessary and sufficient criteria for separability
are still lacking. For other properties, such as distillability, no efficient decision
methods are known, and it is
even difficult to find meaningful
entanglement measures
\cite{Jens+Martin}.
A direction that promises to lead to simpler structures is that of
infinite dimensional subsystems, such as harmonic oscillators or
light modes,
which are commonly denoted as continuous-variable
systems \cite{cvs,cvs2}.
Indeed, for continuous-variable systems the situation
becomes much more transparent if one restricts attention to
Gaussian states (e.g. coherent, squeezed or thermal states) which are, in
any case, the states that are readily experimentally accessible.

Quite recently, it has been realized that it might be a
very fruitful enterprise to apply the
methods from the theory of entanglement
not only to problems of quantum information science, but also
to the study of quantum systems that are typically regarded as
belonging to statistical
physics, systems that consist of a large or infinite number of
coupled subsystems \cite{NielsenPhD,Interacting,exact,Bro}.
Examples of such systems are interacting spin systems, which, like most interacting
systems, exhibit the natural occurrence of entanglement, i.e.,
the ground state is generally an entangled state
\cite{NielsenPhD,Interacting,exact}. It has, furthermore, been
suspected that the study of the entanglement properties of such
systems may shed light on the nature of the structure of classical
and quantum phase transitions \cite{NielsenPhD,exact}. It has
turned out, however, that the theoretical analysis of infinite
spin chains is very complicated and only very rare examples can be
solved analytically. Coupled harmonic oscillator systems allow for
a much better mathematical description of their entanglement
properties than spin systems. Physical realizations of such
systems range from the vibrational degrees of freedom in lattices
to the discrete version of free fields in quantum field theory.
This motivates the approach that we have taken in this work,
namely to investigate the entanglement structure of infinitely
extended harmonic oscillator systems.

In this paper we study a special case, namely a set of harmonic
oscillators arranged on a ring and furnished with a harmonic
nearest-neighbor interaction, i.e., oscillators that are connected
to each other via springs. The paper is organized as follows. In
Section \ref{sec_cov} we provide the basic mathematical tools that are
employed in the analysis following in the remaining sections. We
then move on to derive a simple analytical expression for the
ground state energy of the harmonic oscillator systems. Our main
interest is the computation of entanglement properties of the
ground state of the chain. In Section \ref{sec_logneg} we derive a general
formula for the logarithmic negativity \cite{Nega,Mono} which we
employ as our measure of entanglement. In Section \ref{sec_symm} we present
analytical results that concern the symmetrically bisected chain,
that is the situation where the chain is subdivided into two equal
contiguous parts and the entanglement is calculated between those
parts. We show how to construct a very simple lower bound on the
log-negativity, in the form of a closed-form expression based on
the coupling strengths; that is, no matrix calculations are
necessary. Furthermore, for nearest-neighbor interaction, we show
that the bound is sharp, i.e., gives the exact value of the
log-negativity. Surprisingly, the value of the log-negativity in
this case is independent of the chain length; in particular, it
remains finite. We show in Section \ref{sec_ineq} that the problem is not
reducible to a four-oscillator picture, thereby demonstrating the
non-triviality of the physical system. We then move on to Section
\ref{sec_gen}, where we study general bisections of the chain numerically. We
demonstrate that entanglement is maximized for the
symmetrically bisected chain. Furthermore, and rather counterintuitively,
for asymmetric bisections where one group of oscillators is very
small, and especially when it consists of only one oscillator, we
find that the entanglement decreases if the size of the other
group is increased. We also demonstrate that for large numbers of
oscillators the mean energy of the ground state and the value of
the negativity are proportional and provide an interpretation for
this result. In Section \ref{sec_disc} we discuss our results. We also
provide an intuitive picture that allows to explain the results in
the previous sections.

Generally we have attempted to structure the sometimes somewhat
involved mathematics in such a way, that the reader can skip it
and extract the main physical results easily. We state at the beginning of
each section what main result will be obtained and we
state this result clearly, either in the form of a theorem or at
the end of the section.

\section{Covariance matrix for Gaussian states of the harmonic
chain}\label{sec_cov}

In this section we derive an expression for the covariance matrix of the ground state and of the thermal
states of a set of harmonic oscillators that are coupled via a general interaction that is
quadratic in the position operators (e.g.\ oscillators coupled by springs).
As a byproduct we also give an expression for the energy of the ground state.

Let us first consider the covariance matrix for the ground state
of a single uncoupled harmonic oscillator. The Hamiltonian is
given by (we have adopted units where $\hbar=1$)
\be \hat{H} =
\frac{1}{2m} \hat{P}^2 + \frac{m\omega^2}{2}\hat{X}^2. \ee Denoting
the quadrature operators as a column vector $R$, with
$R_1=\hat{X}$ and $R_2=\hat{P}$, the Hamiltonian can be concisely
rewritten as
\be
    \hat{H} = R^T \twomat{m\omega^2/2}{0}{0}{1/(2m)   } R.
\ee
The covariance matrix $\gamma$ of a general state $\rho$ is given
by \be \gamma_{k,l} = \text{Re }\trace[\rho (R_k - \trace[\rho
R_k]) (R_l - \trace[\rho R_l])]. \ee for $1\leq k,l \leq 2$. For
$\rho_n$ the $n$-th eigenstate of the Hamiltonian,
$\rho_n=|n\rangle\langle n|$, it is a straightforward exercise to
calculate that \be \gamma = (n+1/2)
\twomat{1/(m\omega)}{0}{0}{m\omega}. \ee
 We will
only be interested in the ground state, $\rho_0=
|0\rangle\langle 0|$,
however, since this is the only eigenstate which is Gaussian.

Passing to the harmonic chain consisting of $n$ harmonic oscillators,
we will only consider interactions between the oscillators due to a coupling
between the different position operators.
According to the $(q,p)$-convention we have adopted here,
the vector $R$ of quadrature operators is given by $R_j=\hat{X}_j$ and $R_{n+j}=\hat{P}_j$, for $1\le j\le n$.
The Hamiltonian is then of the form
\be
\hat{H} = R^T \twomat{V m\omega^2/2 }{0}{0}{\identity_{n}/(2m)} R,
\ee
where the $n\times n$-matrix
$V$ contains the coupling coefficients. The Hamiltonian is thus written as a quadratic
form in the quadrature operators; we will call the matrix corresponding to this form the {\em Hamiltonian
matrix} (as opposed to $\hat{H}$, the Hamiltonian {\em operator}).
In the present case, the Hamiltonian matrix is a direct sum of the {\em kinetic matrix} $\identity_n/(2m)$
and the {\em potential matrix} $V m\omega^2/2$.

In this paper, we will consider a harmonic chain ``connected'' end-to-end by a translationally invariant
Hamiltonian. The $V$-matrix of the Hamiltonian is, therefore, a so-called
circulant matrix \cite{HJ2}.
This is a special
case of a Toeplitz matrix
because not only do we have $V_{j,k}=v_{j-k}$, but even $V_{j,k}=v_{(j-k) \bmod n}$
for $1\leq j,k\leq n$,
due to the end-to-end connection.
We can easily write the coefficients $v_k$ in terms of the coupling coefficients.
For a nearest-neighbor coupling with ``spring constant'' $K$, the potential term of the Hamiltonian reads
\be
\sum_{k=1}^{n}
\frac{m\omega^2}{2}\hat{X}_k^2 + K(\hat{X}_{(k+1)\bmod n}-\hat{X}_k)^2.
\ee
Therefore, we have \be v_0 =
1+4K/(m\omega^2),\,\,\,v_1=-2K/(m\omega^2).\ee
More generally, including $k$-th nearest-neighbor couplings with spring constants $K_k$, and defining
\be
\alpha_k = \frac{2 K_k}{m\omega^2},
\ee
we have
\beas
v_0 &=& 1+2(\alpha_1+\alpha_2+\ldots), \\
v_j &=& -\alpha_j, \text{ for }j>0.
\eeas

The calculation of the corresponding covariance matrix can now
proceed via a diagonalisation of the Hamiltonian matrix, which
effectively results in a decoupling of oscillators. Since the
commutation relations between the quadrature operators must be
preserved, the diagonalisation must be based on a {\em symplectic
transformation} $S\in Sp(2n,\R)$. This means that we can
only use equivalence transformations $C\mapsto C'=S^TCS$ such that
$S^T \Sigma S=\Sigma$, where, in the $(q,p)$-convention, the {\em
symplectic matrix} $\Sigma$ is given by
\be
\Sigma=\twomat{0}{\identity_n}{-\identity_n}{0}.
\ee
This real skew-symmetric
matrix incorporates the canonical commutation relations
between the canonical coordinates.
Fortunately, because the kinetic
matrix is a multiple of the identity, the Hamiltonian matrix
can be diagonalized by an {\em orthogonal} equivalence of the form
\be
C\longmapsto C'=(S\oplus S)^T C (S\oplus S),\ee where $S$ is the
real orthogonal $n\times n$-matrix that diagonalizes the potential
matrix $V$. It is readily checked that the resulting
transformation is indeed a symplectic one. In fact, $S\oplus S$ is
an element of the maximal compact subgroup of
$Sp(2n,\R)$.

So $C'$ is now diagonal and of the form
$C'=(m\omega^2/2)V' \oplus \identity_{n}/(2m) $, 
where $V'$ is the
diagonal $n\times n$-matrix with entries $\eta_j$, $1\leq j\leq n$,
the eigenvalues of $V$.
The covariance matrix $\gamma'$ of the ground state of the transformed Hamiltonian consists therefore just of
single-oscillator covariance matrices with parameter $\omega_j=\omega\sqrt{\eta_j}$,
$1\leq j\leq n$,
and is diagonal itself, to wit,
\beas
\gamma' &=& (\gamma'_x\oplus\gamma'_p)/2, \\
(\gamma'_x)_{j,j} &=& 1/(m\omega_j), \\
(\gamma'_p)_{j,j} &=& m\omega_j.
\eeas
The covariance matrix $\gamma$ in the original
coordinates is then obtained by transforming $\gamma'$ back,
\beas
\gamma &=& (S\oplus S)\gamma' (S\oplus S)^T \\
&=&  [(S\gamma'_x S^T) \oplus (S\gamma'_p S^T)] /2\\
&=&  [(V^{-1/2}/(m\omega))\oplus(m\omega V^{1/2})]/2.
\eeas
To simplify the notation,
we will henceforth set $m=1$ and $\omega=1$.
So we have a simple formula for the covariance matrix in terms of the potential matrix $V$,
\beas
\gamma &=&  (\gamma_x\oplus\gamma_p)/2, \\
\gamma_x &=& V^{-1/2}, \\
\gamma_p &=& V^{1/2}.
\eeas

Using this same derivation, we can also easily find a formula for the energy of the ground state.
We will need this result in Section \ref{sec_gen}, where we will compare the log-negativity of a state
to its energy.
Indeed, the ground state energy of a single oscillator is $\hbar\omega/2$. In the decoupled description
of the ground state of the chain, the oscillators have energy
$\hbar\omega\sqrt{\eta_j}/2$, with $\eta_j$, $1\leq j\leq n$, being
the eigenvalues of the potential matrix $V$. The total
ground state energy is $E=(\hbar\omega/2)\sum_{j=1}^{n}
\sqrt{\eta_j}$. Denoting $\hbar\omega/2$ by $E_0$, we
therefore have
\be
E = E_0 \trace [V^{1/2}].
\ee
Finally, we turn to Gibbs
states corresponding to some temperature $T>0$,
the states associated with the canonical ensemble, given by
\be\rho(\beta)=\exp(-\beta \hat H)/\trace[ \exp(-\beta \hat H)],\ee
where
$\beta= 1/T$.
Again, one can obtain the
covariance matrix $\gamma(\beta)$ of the state $\rho(\beta)$
in a convenient manner in the
basis in which the Hamiltonian matrix is diagonal.
The
$2n\times 2n$ diagonal matrix $\gamma'(\beta)$
can be obtained using the virial theorem: the
mean potential energy and the kinetic energy of
a single oscillator are identical and half the
mean energy of the system at inverse temperature $\beta$.
Using this procedure one obtains
\beas
\gamma'(\beta) &=& (\gamma'_x(\beta)\oplus\gamma'_p(\beta))/2, \\
(\gamma'_x(\beta))_{j,j}  &=& \frac{1}{m\omega_j}\left(
1+ \frac{2}{\exp(\beta \omega_{j}) -1}
\right), \\
(\gamma''_p(\beta))_{j,j} &=& m\omega_j
\left(
1+ \frac{2}{\exp(\beta \omega_{j}) -1}
\right).
\eeas
In the convention where $m=1$, $\omega=1$, one gets
\beas
\gamma(\beta) &=& (\gamma_{x}(\beta)\oplus \gamma_{p}(\beta) )/2,\\
\gamma_x(\beta) &=& V^{-1/2} \Bigl(\identity_{n} + 2(\exp( \beta
V^{1/2}) -\identity_{n})^{-1} \Bigr), \\
\gamma_p(\beta) &=& V^{1/2}  \Bigl(\identity_{n} + 2(\exp( \beta
V^{1/2}) -\identity_{n})^{-1} \Bigr),
\eeas
for the covariance matrix of a Gibbs state in the
original canonical coordinates.

\section{General formula for the logarithmic negativity}\label{sec_logneg}
In this section we derive a general formula for the logarithmic negativity of a
Gaussian state of $n$ coupled harmonic oscillators with respect to a bipartite split,
given the covariance matrix $\gamma$ of the
Gaussian state. This set may consist of all $n$ oscillators or
of a subset of $m<n$ oscillators.
The only restriction is that the covariance matrix must be a direct sum
of a position part $\gamma_x$ and a momentum part $\gamma_p$,
i.e., there must be no correlations between positions and momenta.
The resulting formula can be found at the end of this section.

Let $m_1$ and $m_2$ be the sizes of the two groups of oscillators the entanglement between which we wish
to calculate, and let $m=m_1+m_2\leq n$. From Section \ref{sec_cov}, we know that
the covariance matrix $\gamma$ of the ground state of the harmonic chain
is given by $\gamma = (\gamma_x\oplus \gamma_p)/2$,
where $\gamma_x=V^{-1/2}$ and $\gamma_p=V^{1/2}$.
In order to calculate the entanglement between two disjoint groups of oscillators in this state,
we need to consider the covariance matrix associated with the reduced
state of the $m$ oscillators of the two groups. This covariance
matrix $\mu$ -- from now on also referred to as reduced covariance matrix --
is given by the $2m\times 2m$
principal submatrix of $\gamma$ that consists of those
rows and colums of $\gamma$ that correspond to
the canonical coordinates of either group $1$ or group $2$.
If $m=n$, meaning that the whole set of $n$ oscillators
is considered, this step is not necessary. The reduced covariance
matrix $\mu$ is again of the form
\be\mu=(\mu_{x}\oplus \mu_{p})/2,\ee
where both $\mu_{x}$ and $\mu_{p}$ are $m\times m$-matrices.

Taking the {\em partial transpose}
of a covariance matrix
corresponds to changing the sign of the momentum variables of the
oscillators in the second group. This operation maps the covariance matrix
$\mu$ to \be\mu^\Gamma=P\mu P\ee with
\be
P=P_x \oplus P_p,\,\,\,\,P_x=\identity_m;
\ee
$P_p$ is a $m\times m$ diagonal matrix. Specifically,
the $j$-th diagonal element of $P_p$ is $1$ or $-1$, depending on whether
the oscillator on position $1\leq j\leq m$ belongs
to group 1 or 2, respectively.

The {\em logarithmic negativity} \cite{Nega,Mono} of a state
is defined as the logarithm of the
trace norm of the partial transpose of the state. The
negativity is an entanglement measure in the sense that it is a
functional that is monotone under local
quantum operators \cite{Mono,PhD}. To date it is the
only feasible measure of entanglement for mixed
Gaussian quantum states.
The definition of the logarithmic negativity
can be easily translated into
an expression which does not involve the
state itself, but rather the covariance matrix
of the state: as the trace norm is unitarily invariant,
one has the freedom to choose a basis for which the
evaluation of the trace norm becomes particularly
simple. More specifically, one may make
use of the {\em Williamson normal form} \cite{Wil}
for the partial transpose of the covariance matrix.
The problem of evaluating the logarithmic
negativity is then essentially reduced to
a single-mode problem. This procedure gives
rise to the formula \cite{Mono}
\be
N = -\sum_{k=1}^{2m}
\log_2(\min(1,2|{\lambda}_k(i \Sigma^{-1}\mu^\Gamma)|)),
\ee
where ${\lambda}_k(i \Sigma^{-1} \mu^\Gamma) $,
$1\leq k\leq 2m$, are the eigenvalues of
$i\Sigma^{-1}\mu^{\Gamma}$.
$\Sigma$ is the symplectic matrix
\be
\Sigma=\twomat{0}{\identity_m}{-\identity_m}{0}.
\ee
Since $\Sigma^{-1} = -\Sigma$,
we have to calculate the spectrum of the matrix $B=-i \Sigma P \mu P$,
giving $2m$ real eigenvalues $\lambda_k(B)$ of $B$.
Then the logarithmic negativity equals
$N = -\sum_{k=1}^{2m}
\log_2\min(1,2|\lambda_k(B)|)$.
This formula can be further simplified due to
the direct sum structure of $\mu=(\mu_{x}\oplus
\mu_{p})/2$. Simplification of $B$ yields
\be
B = \frac{i}{2}\twomat{0}{-P_p \mu_p P_p}{\mu_{x}}{0}.
\ee
The eigenvalue equation of a block matrix of this form reads
\be
\twomat{0}{X}{Y}{0} \twovec{u}{v} = \lambda \twovec{u}{v},
\ee
which is equivalent to the coupled system of equations $Xv=\lambda u$ and $Yu=\lambda v$.
Substituting one equation in the other yields $XYu=\lambda^2 u$, hence the eigenvalues of the block matrix
are plus and minus the square roots of the eigenvalues of $XY$. In particular, the eigenvalues of $B$ are
$\pm (\lambda_j(\mu_x P_p \mu_p P_p/4 ))^{1/2}$, $1\leq j\leq
m$.
Because of the $\pm$-sign,
taking the absolute value of the eigenvalues has the effect of doubling
the eigenvalue multiplicity. Hence,
\be
N = -\sum_{j=1}^{m}
\log_2\min(1,\lambda_j(\mu_x P_p \mu_p P_p)),
\ee
which is finally the resulting formula
of the logarithmic negativity in terms of the
matrices $\mu_{x}$ and $\mu_{p}$.

\section{The Symmetrically Bisected Harmonic Chain} \label{sec_symm}
\begin{figure}
\leavevmode
\epsfxsize=6cm
\centerline{
\epsfbox{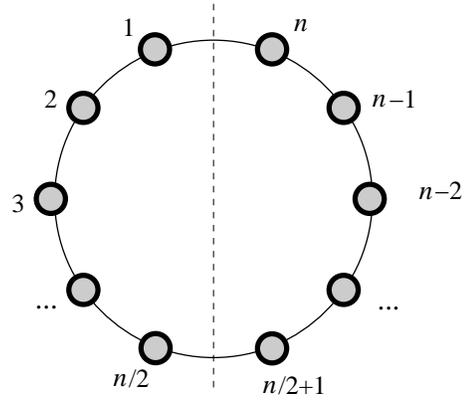}}
\smallskip
\caption{The symmetrically bisected harmonic chain. The oscillators
$1$ to $n/2$ form group $1$, the oscillators $n/2+1$ to $n$ form group $2$.}
\label{fig_plot1a}
\end{figure}

In this section we present exact analytical results for the log-negativity in a chain
of $n$ harmonic oscillators with a translationally invariant coupling.
Moreover, we shall be interested here in the most symmetric case of calculating the entanglement with
respect to a symmetric bisection of the chain.
That is, the number $n$ of oscillators should be even and
the oscillators in positions 1 to $n/2$ constitute group 1, the others group 2
(see Figure \ref{fig_plot1a}).
Hence, in the notation of Section \ref{sec_logneg} $\mu=\gamma$.

Using the result of Section \ref{sec_logneg},
we find that the logarithmic negativity of a
symmetrically bisected oscillator chain, of length $n$ and with potential matrix $V$, is equal to
\be
N =-\sum_{j=1}^{n}
\log_2(\min(1,\lambda_j(Q))),
\ee
with
\beas
Q&=& V^{-1/2}P V^{1/2}P \\
P&=& \identity_{n/2}\oplus(-\identity_{n/2}).
\eeas

\subsection{Symmetry properties of $Q$}
We begin our analytical investigations by studying a more general object than $Q$,
namely the matrix
\be
R = G^{-1}PGP.
\ee
Here, $G$ is a general real {\em circulant} matrix that is symmetric under transposition.
Since $G$ is symmetric and circulant, it can be written in $2\times 2$ block form as
\be G=\twomat{G'}{G''}{G''}{G'}.\ee
We will first show that $R$ exhibits the same block structure.
Define the $n\times n$ {\em flip-matrix} $F=F_n$ as
\be
F_{i,j}=\delta_{i,n+1-j}.
\ee
To simplify the notation, we will mostly refrain from mentioning the size $n$ of $F$;
the mathematical context should make it clear which $n$ is being used.
\begin{lemma}\label{th_lem1}
The matrix $R$ can be written in $2\times 2$ block form as \be R = \twomat{A}{B}{B}{A}.\ee
\end{lemma}
{\em Proof.}
Since $G$ is circulant and symmetric, $FGF=G$, which
is true for every symmetric Toeplitz matrix.
Also, $FPF=-P$ holds.
Writing \be R=\twomat{A}{B}{C}{D},\ee symmetry demands
that $C=FBF$ and $D=FAF$.
Furthermore, both $G$ and $P$ are also invariant under the $F_{n/2}\oplus F_{n/2}$ symmetry.
Hence, $R$ exhibits this symmetry too, i.e.,  $(F\oplus F)R(F\oplus F)=R$. Thus, $A$ and $B$ are
invariant under $F$.
\qed

Matrices with this block structure can be brought in block diagonal form
using a similarity transform:
\begin{lemma}\label{th_lem2}
Let ${\cal S} = (P+F)/\sqrt{2}$. Then
\be
{\cal S}\twomat{A}{B}{B}{A}{\cal S} = (A+BF)\oplus(A-BF).
\ee
\end{lemma}
{\em Proof.} This follows by direct calculation and noting that
${\cal S}^{-1}={\cal S}$ and
\be{\cal S}=\twomat{\identity_n}{F}{F}{-\identity_n}/\sqrt{2}.\ee
\qed

We now specialize the above results for $R$ to the matrix
\be
Q=V^{-p}P V^pP,
\ee
with $V$ again a general real symmetric circulant matrix.
The power $p$ remains hitherto unspecified.
Note that any power of a symmetric circulant matrix is again symmetric circulant.
Again, the $n\times n$-matrix
$V$ can be written in $2\times2$ block form
\be\twomat{V'}{V''}{V''}{V'},\ee
where both $V'$ and $V''F$ are Hermitian.
By Lemma \ref{th_lem1}, $Q$ can similarly be written as the block matrix
\be\twomat{Q'}{Q''}{Q''}{Q'}.\ee
The following Lemma is crucial for the rest of the calculations.
\begin{lemma}\label{th_lem3}
With the previous notations,
$Q'+Q''F = (Q'-Q''F)^{-1}$ and
\beas
\det(Q'+Q''F) &=& \exp(-p\trace [F\log(V)]), \\
\det(Q'-Q''F) &=& \exp(+p\trace [F\log(V)]).
\eeas
For $p$ in the interval $0\le p\le 1$, the following also holds:
if $V''F\ge 0$, then $Q'+Q''F\le\identity_{n/2}$, and
if $V''F\le 0$, then $Q'-Q''F\le\identity_{n/2}$.
\end{lemma}
{\em Proof.}
Consider ${\cal S}Q{\cal S}$.
On one hand, we have, by Lemma \ref{th_lem2},
\be
{\cal S}Q{\cal S} = (Q'+Q''F)\oplus (Q'-Q''F),
\ee
and similarly,
${\cal S}V{\cal S} = (V'+V''F)\oplus (V'-V''F)$.
Also,
${\cal S}P{\cal S} = F$, as a short calculation shows.
On the other hand, we also have
\beas
{\cal S}Q{\cal S} &=& {\cal S} V^{-p}PV^pP {\cal S} \\
&=& {\cal S} V^{-p}{\cal S}{\cal S}P{\cal S}{\cal S}V^p{\cal S}{\cal S}P {\cal S}\\
&=& ({\cal S} V{\cal S})^{-p} {\cal S}P{\cal S} ({\cal S}V{\cal S})^p
{\cal S}P {\cal S},\nonumber
\eeas
and therefore
\beas
{\cal S}Q{\cal S}
&=& ((V'+V''F)^{-p}\oplus (V'-V''F)^{-p}) F \times \\ && ((V'+V''F)^p\oplus (V'-V''F)^p) F \\
&=& ((V'+V''F)^{-p}\oplus (V'-V''F)^{-p}) \times \\ && ((V'-V''F)^p\oplus (V'+V''F)^p) \\
&=& (V'+V''F)^{-p}(V'-V''F)^p \oplus \\ && (V'-V''F)^{-p}(V'+V''F)^p.
\eeas
Identifying the blocks in the two expressions for ${\cal S}Q{\cal S}$,
we get
\beas
Q'+Q''F &=& (V'+V''F)^{-p}(V'-V''F)^p, \\
Q'-Q''F &=& (V'-V''F)^{-p}(V'+V''F)^p,
\eeas
so that $Q'+Q''F$ is the inverse of $Q'-Q''F$ and
\be
\det(Q'+Q''F) = \left(\frac{\det(V'-V''F)}{\det(V'+V''F)}\right)^p.
\ee
Furthermore,
$\log V={\cal S}(\log(V'+V''F)\oplus\log(V'-V''F)){\cal S}$,
hence
\beas
&& \trace [F\log V] \\
&&= \trace[{\cal S}F{\cal S}(\log(V'+V''F)\oplus\log(V'-V''F))] \\
&&= \trace [P (\log(V'+V''F)\oplus\log(V'-V''F))] \\
&&= \trace [\log(V'+V''F)-\log(V'-V''F)] \\
&&= \log\frac{\det(V'+V''F)}{\det(V'-V''F)}.
\eeas
This then yields
\beas
\det(Q'+Q''F) &=& \exp(-p\trace [F\log V]), \\
\det(Q'-Q''F) &=& \exp(+p\trace [F\log V]).
\eeas

Considering the second assertion, if $V''F\ge0$, then
$V'+V''F\ge V'-V''F$, and, by L\"owner's theorem \cite{HJ2},
\be(V'+V''F)^p\ge (V'-V''F)^p\ee
for $0\le p\le 1$.
Hence, for any vector $x\neq 0$ satisfying an equation
$(V'-V''F)^p x=\lambda(V'+V''F)^p x$, it follows that $\lambda$ must be less than or equal
to 1 (to see this, take the inner product of both sides with the vector $x$).
Rearranging the equation to \be(V'+V''F)^{-p}(V'-V''F)^p x=\lambda x,\ee
which is just $(Q'+Q''F) x=\lambda x$, yields that
the $\lambda$ for which such an $x$ exists are precisely the eigenvalues of $Q'+Q''F$.
Hence, under the condition $V''F\ge 0$, the eigenvalues of $Q'+Q''F$ are less than or equal to 1.
Similarly, if $V''F\le0$, we proceed in an identical way to show that the eigenvalues of $Q'-Q''F$
are less than or equal to 1.
\qed

\subsection{A lower bound on the negativity}
We will now apply Lemma \ref{th_lem3} to the case $p=1/2$ and $V$ being
the potential matrix of the oscillator chain
to obtain a lower bound on the logarithmic negativity:
\begin{theorem}\label{th_1}
The logarithmic negativity of the bisected oscillator chain of
length $n$ obeys
\be
N \ge|\trace[F\log_2(V)]|/2.
\ee
If $V''F$ is semidefinite (i.e., either positive or negative semidefinite), then equality holds.
\end{theorem}
{\em Proof.}
To calculate the negativity we need the eigenvalues of $Q$ with $p=1/2$ that are smaller than 1.
By Lemma \ref{th_lem2}, the spectrum of $Q$ is the union of the spectra of $Q'+Q''F$ and of $Q'-Q''F$.
By Lemma \ref{th_lem3}, for $V$ matrices satisfying the $V''F\ge0$ condition,
the eigenvalues of $Q$ smaller than 1 are the
eigenvalues of $Q'+Q''F$. Furthermore,
\beas
\trace [\log(Q'+Q''F)]&=&\log\det(Q'+Q''F)\\
&=&-p\trace [F\log(V)].
\eeas
Setting $p=1/2$ then gives $N
=\trace [F\log_2(V)]/2$.
On the other hand, if $V''F\le0$, it is the eigenvalues of
$Q'-Q''F$ that we need to consider. Since $\trace[\log(Q'-Q''F)] =
+p\trace[F\log(V)]$, we find
\be N = -\trace [F\log_2(V)]/2.\ee
For general $V''F$, we first note that the general formula for the negativity can be written as
\be N = -\trace[\log_2\min(\identity_n,Q)].\ee
For commuting $X$ and $Y$, $\min(X,Y)$ is the elementwise minimum
in the eigenbasis of $X$ (and $Y$).
By Lemma \ref{th_lem2}, we then have $N=-\trace[\log_2\min(\identity_n,(Q'+Q''F)\oplus(Q'-Q''F))]$
and this is also equal to
$N=-\trace[\log_2\min(\identity_{n/2},Q'+Q''F)]-\trace[\log_2\min(\identity_{n/2},Q'-Q''F)]$. From Lemma
\ref{th_lem3}
we also know that $Q'+Q''F$ and $Q'-Q''F$ are each other's inverse. Hence,
\beas
N=-\trace[\log_2\min(Q'+Q''F,Q'-Q''F)],
\eeas
and, because the two arguments of min commute,
$N=-\trace[\min(\log_2(Q'+Q''F),\log_2(Q'-Q''F))]$.
Finally, the trace of a minimum is smaller than or equal to
the minimum of the traces, so that
\beas
N\ge
\max(\trace[\log_2(Q'+Q''F)], \trace[\log_2(Q'-Q''F))].
\eeas
Because the two arguments of max are each other's negative,
the maximum amounts to taking the absolute
value of, say, the first argument. Hence,
$N\ge |\trace[\log_2(Q'+Q''F)]| = |\trace[F\log_2(V)]|/2$, where the last equality follows from the
first part of the proof.
\qed

For the nearest-neighbor Hamiltonian, $V$ is of the form
\be
V=\left(
\begin{array}{cccccc}
v_0&v_1&0&\cdots&0&v_1 \\
v_1&v_0&v_1&      & &0 \\
0&v_1&v_0&   \ddots  &   & \vdots  \\
\vdots& & \ddots &\ddots & v_{1} & 0\\
0&&&v_{1}&v_{0}&v_{1}\\
v_1&0&\cdots&0&v_1&v_0
\end{array}
\right),
\ee
so
\begin{eqnarray}
V' &=&\left(
\begin{array}{ccccc}
v_0&v_{1}&0& \cdots&0 \\
v_{1}& v_{0}& v_{1}&   &  \vdots \\
0 &\ddots &\ddots &\ddots & 0\\
\vdots &&v_{1} &v_{0}& v_{1}\\
0 &\cdots & 0 &v_{1}&v_0
\end{array}
\right), \\[0.2cm]
V''F &=& \left(
\begin{array}{ccccc}
v_1    & 0    & \cdots & &0 \\
0      & 0    &        & & \\
\vdots &      & \ddots & &\vdots \\
       &      &       &0&  0\\
0      &\cdots&       & 0&v_1
\end{array}
\right).
\end{eqnarray}
As $v_1\le 0$, $V''F$
is obviously a (negative) semidefinite matrix. Therefore,
the nearest-neighbor Hamiltonian satisfies the equality condition of the Theorem, and
the logarithmic negativity of the bisected harmonic chain with nearest-neighbor Hamiltonian equals
$N =|\trace[F\log_2(V)]|/2$.

\subsection{An explicit formula in the coupling}
The bound of Theorem \ref{th_1} is actually a very simple one because we can give an explicit formula
for $|\trace[F\log_2(V)]|$ in terms of the coupling coefficients
$\alpha_j$, as follows.
\begin{theorem}\label{th_2}
For a translationally invariant potential matrix
$V$ with coupling coefficients $\alpha_1$, $\alpha_2$, \ldots $\alpha_m$,
\be
|\trace[F\log_2(V)]| = \log_2(1+4(\alpha_1+\alpha_3+\ldots)).
\ee
\end{theorem}
Note, in this formula, the absence of the coefficients with even index.

{\em Proof.}
The eigenvalue decomposition of a general circulant $n\times n$ matrix $V$ is
very simple to calculate. For convenience of notation,
we use matrix indices starting from zero instead of 1.
Let $V_{k,l}=v_{k-l}$, then
$V=\Omega^\dagger\Lambda\Omega$, with $\Omega$ the kernel matrix of the discrete
Fourier transform,
\be
\Omega_{k,l}=\exp\bigl(kl \frac{2\pi i}{n}\bigr)/\sqrt{n},
\ee
with $0\le k,l\le n-1$.
This matrix is unitary and symmetric.
The   eigenvalues $\Lambda_k$ are related to
$v_l$ via a discrete Fourier transform according to
\be
\Lambda_k = \sum_{l=0}^{n-1} \exp\left(\frac{2\pi i}{n}k l\right) v_l.
\ee
For real symmetric $V$ this gives
\beas
\Lambda_k = v_0+2v_1\cos\bigl(k \frac{2\pi}{n}\bigr)+2v_2\cos\bigl(2k
\frac{2\pi}{n}\bigr)+\cdots .
\eeas
It is now a straightforward calculation to obtain an expression for
$\trace[F\log_2(V)]$.
First,
\beas
&& (\Omega F\Omega^\dagger)_{k,l} \\
&&= \sum_{j,j'=0}^{n-1} \exp(jk\frac{2\pi i}{n}) \delta_{j,n-1-j'}
\exp(-j'l\frac{2\pi i}{n})/n  \\
&&= \sum_{j=0}^{n-1}  \exp([jk-(n-1-j)l]\frac{2\pi i}{n})/n \\
&&= \sum_{j=0}^{n-1}  \exp([j(k+l)-(n-1)l)]\frac{2\pi i}{n})/n.
\eeas
All elements are zero except those for which $k+l$ is an integer multiple of $n$, i.e.,  either $k=l=0$ or $k+l=n$:
\be
(\Omega F\Omega^\dagger)_{0,0}
= \sum_{j=0}^{n-1}\exp((0j+0)\frac{2\pi i}{n})/n
= 1
\ee
and
\beas
(\Omega F\Omega^\dagger)_{n-l,l}
&=& \sum_{j=0}^{n-1}\exp((jn-(n-1)l)\frac{2\pi i}{n})/n  \\
&=& \exp(-l\frac{2\pi i}{n}).
\eeas
In the calculation of $\trace [F\log_2(V)]$ we only need the non-zero
diagonal elements of $\Omega F\Omega^\dagger$, which are the $(0,0)$ and the $(n/2,n/2)$ elements. Hence
\beas
&& \trace [F\log_2(V)] \\
&&=\log_2(\Lambda_{0}) +
\log_2(\Lambda_{n/2}) \exp((n/2)\frac{2\pi i}{n}) \\
&&= \log_2\frac{v_0+2v_1+2v_2+\cdots}{v_0-2v_1+2v_2-\cdots}.
\eeas
Inserting the relations between the elements of $V$ and the coupling coefficients $\alpha_j$
\beas
v_0 &=& 1+2(\alpha_1+\alpha_2+\ldots), \\
v_j &=& -\alpha_j,\text{ for }j>0,
\eeas
yields the stated formula.
\qed

For the nearest-neighbor Hamiltonian, the only non-zero $\alpha_j$ coefficient
is $\alpha\equiv \alpha_1$, giving rise to the following simple expression for
the logarithmic negativity.
\begin{corollary}\label{th_cor1}
For the nearest-neighbor Hamiltonian with coupling
    coefficient $\alpha\geq 0$,
    the
    logarithmic negativity of the bisected chain of length $n$
    is given by
\be
N =\frac{1}{2}\log_2(1+4\alpha).
\ee
\end{corollary}
It is remarkable indeed
that the negativity is independent of $n$, the chain length.

\subsection{Other potential matrices}
To conclude this section, we will prove
that any other circulant symmetric
potential matrix does not satisfy the equality condition of Theorem
\ref{th_2} so that
the negativity will in general be larger than the lower bound and, moreover,
dependent on the size $n$ of the chain.
Consider first a Hamiltonian with a nearest-neighbor coupling of strength $\alpha_1$ and
a next-nearest-neighbor coupling of strength $\alpha_2$, where
$\alpha_1,\alpha_2>0$.
The matrix $V''F$ is then of the form
\be
V''F=\left(
\begin{array}{cccc}
-\alpha_1&-\alpha_2&\cdots&0 \\
-\alpha_2&0  &      &  \\
\vdots&&&\\
&&0&-\alpha_2\\
&\cdots&-\alpha_2&-\alpha_1
\end{array}
\right).
\ee
The non-zero eigenvalues of this matrix are those of the submatrix
\be\twomat{-\alpha_1}{-\alpha_2}{-\alpha_2}{0}.\ee
As its determinant is negative, $-\alpha_2^2$, it is not a
definite matrix, hence neither is $V''F$.

Generally, a $k$-th neighbor coupling $\alpha_k$,
i.e.,  a coupling between oscillators $k$ places apart,
yields a matrix $V''F$ which is {\em Hankel} \cite{HJ2}
and has $-\alpha_k$ on two skew-diagonals.
If there is a $k$ such that $\alpha_k$ and $\alpha_{k+1}$ are non-zero but $\alpha_{k+2}=0$,
then $V''F$ contains a $2\times 2$ principal submatrix of the form
\be\twomat{-\alpha_{k}}{-\alpha_{k+1}}{-\alpha_{k+1}}{0},\ee
which is again not definite. Hence, in
that case, $V''F$ is not semidefinite. Now, if one fixes the interactions and then let
$n$ grow (which is exactly the setting here),
there will always be some point when $V''F$ will exhibit a zero skew-diagonal and, hence,
is not semidefinite.
\section{Inequivalence to a four-oscillator problem} \label{sec_ineq}

\begin{figure}
\leavevmode
\epsfxsize=6cm
\centerline{
\epsfbox{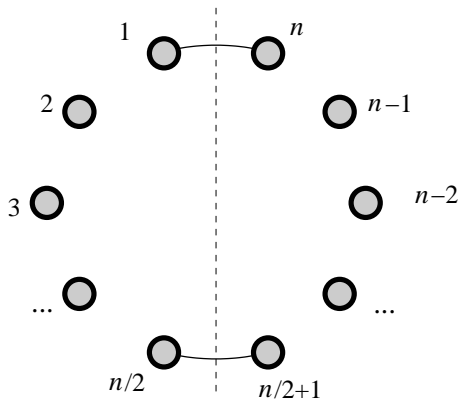}}
\smallskip
\caption{There exist no symplectic transformations that
decouple all but four oscillators from each other in the
case of the bisected harmonic chain with nearest-neighbor
Hamiltonian.}
\label{fig_plot1b}
\end{figure}

At this point, one might be tempted to think that the
independence of the log-negativity of the chain length $n$,
in the case of nearest-neighbor interaction,
is a consequence of the presumption that the bisected harmonic chain
of length $n\geq 4$
with nearest-neighbor interaction is in fact equivalent
to a much simpler problem: there could be an appropriate
choice of basis of the Hilbert spaces
of system 1 and 2, corresponding to a symplectic
transformation, such that, in effect, only
those four oscillators that are adjacent to the split boundary
would be in an entangled state. The other $n/2-2$
oscillators of each system would then be
in pure product states, thereby not contributing to
the logarithmic negativity. This would mean that one
could locally disentangle all but four oscillators with
local symplectic transformations (see Figure \ref{fig_plot1b}).

If this indeed were the case, then symplectic transformations $S_{1}, S_{2}\in
Sp(n,\R)$ would exist such that
\beas
\gamma &=&
(S_{1}\oplus S_{2})^{T} \gamma' (S_{1}\oplus S_{2}) \\
\gamma' &=& (\identity_{n-4}/2) \oplus \gamma_{12}\oplus \tilde
\gamma_{12} \oplus (\identity_{n-4}/2)
\eeas
(note that we are using a quadrature ordering convention here that
is different from the one used in the rest of this paper).
Here, $\gamma_{12}$ and $\tilde\gamma_{12}$ are
$4\times 4$-covariance matrices
associated with the oscillators $1$ and $n$ on the one hand and
$n/2$ and $n/2+1$ on the other hand,
and $\identity_{n-4}/2$ is the covariance matrix of
the pure product states of the remaining $n/2-2$
oscillators of system 1 and 2, respectively.
If for any $n$ such a basis change could be performed,
leading to the same covariance matrices
$\gamma_{12}$ and $\tilde\gamma_{12}$,
then the invariance of the logarithmic
negativity of the bisected chain -- the statement of
Corollary \ref{th_cor1} -- would follow as
a trivial consequence. We will briefly show, however,
that this is not the case.

Consider the eigenvalues
of $B' = i \Sigma^{-1} \gamma'^{\Gamma}$
\be
B' =i \Sigma^{-1} P \Bigl(
(\identity_{n-4}/2) \oplus \gamma_{12}\oplus \tilde\gamma_{12}\oplus
(\identity_{n-4}/2)
\Bigr) P.
\ee
The spectrum of the corresponding matrix
$Q'$ that enters in the formula for the negativity
can easily be evaluated using the procedure mentioned in Section \ref{sec_logneg}.
It is given by
\be\sigma(Q')=\{1,...,1,q_1, q_2, q_3, q_4\},\ee
where $q_1,\dots,q_4 >0$, and $1$ appears $n-4$ times.

\begin{figure}
\leavevmode
\epsfxsize=8.5cm
\centerline{
\epsfbox{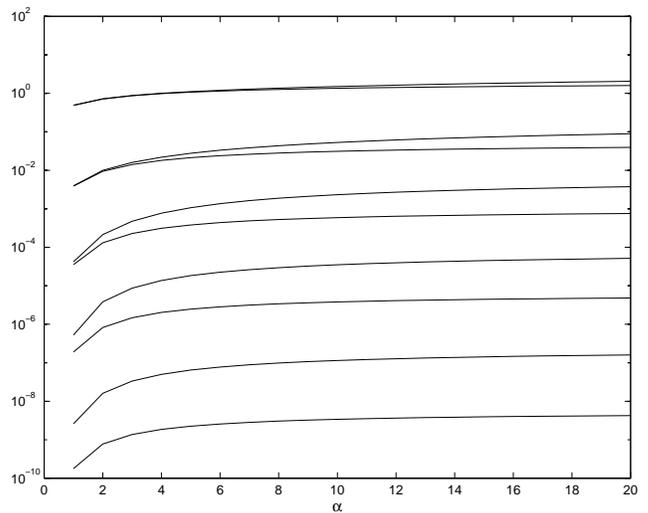}}
\smallskip
\caption{Positive eigenvalues of $Q-\identity$ versus $\alpha$, for a chain of size $n=20$.
From Lemma \protect\ref{th_lem3} it follows that the eigenvalues
of $Q$ come in reciprocal pairs; hence,
the plot shows that in this case all eigenvalues of $Q$ are either larger than 1 or smaller than
1. For other chain sizes, the eigenvalues behave in a similar way.
This shows that the symmetrically bisected chain typically cannot be reduced to the system depicted
in Figure \protect\ref{fig_plot1b}.}
\label{fig_plot9}
\end{figure}

Now we can confront this result with the spectrum of the matrix
$Q$ of the harmonic chain as is. A simple numerical calculation
yields the values depicted in Figure \ref{fig_plot9}.
Since the eigenvalues of $Q$ come in
reciprocal pairs, we only show the eigenvalues larger than 1;
furthermore, we subtract 1 from them and show the result on a
logarithmic scale (in order to clearly distinguish all
eigenvalues). In the case depicted, $n=20$, we see that 10
eigenvalues are larger than 1, for any value of the coupling
constant $\alpha$. Furthermore, the 10 remaining eigenvalues are all
smaller than 1.
This means that, in fact, $1$ is not included in the spectrum of $Q$, which is
completely at variance with the result for $Q'$ of the
purported reduced chain. Hence,
we arrive at the statement that not even a single oscillator
can be exactly decoupled from all the others by the application on an
appropriate local symplectic transformation.

This analysis shows that the coupled bisected chain with
nearest-neighbor interaction can not be reduced to a
problem of only two pairs of interacting oscillators.
In Section \ref{sec_disc} -- equipped with further
results from numerical
investigations -- we will discuss these findings and
present an intuitive picture of the correlations present
in the ground state of this system of coupled oscillators.

\section{General bisections}\label{sec_gen}
In this section we turn towards more general problems, exhibiting less symmetry. As these
problems are much more difficult to solve analytically, we basically have restricted ourselves to
numerical calculations and we only give analytical results for small subproblems, valid in some
asymptotic regime only.

\subsection{Asymmetrical bisections}
\begin{figure}
\leavevmode
\epsfxsize=8cm
\epsfbox{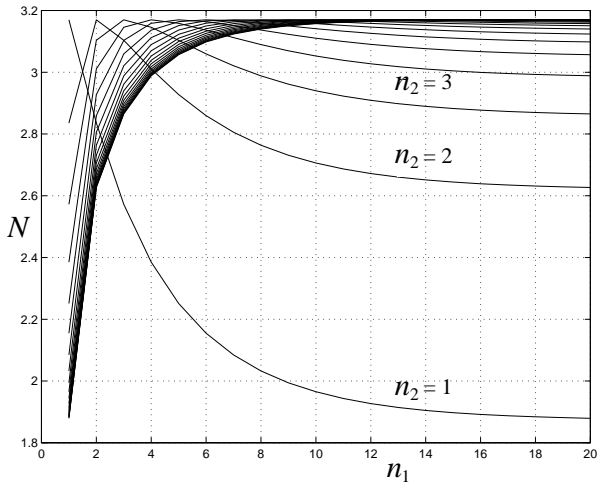}
\caption{Logarithmic negativity $N$ of a harmonic chain bisected in groups of size $n_1$ and $n_2$.
The interaction is nearest-neighbor with coupling $\alpha=20$.}
\label{fig_plot1}
\end{figure}

\begin{figure}
\leavevmode
\epsfxsize=8cm
\epsfbox{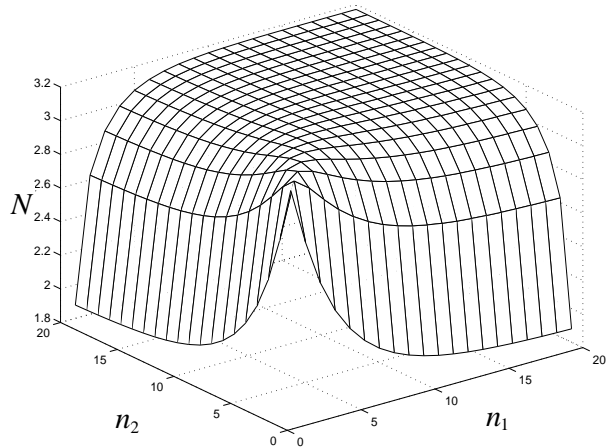}
\caption{Same as Figure \protect\ref{fig_plot1}, but seen from a different viewpoint.}
\label{fig_plot2}
\end{figure}

In Figures \ref{fig_plot1} and \ref{fig_plot2}
we show the results of a numerical calculation for asymmetrically bisected
chains with nearest-neighbor coupling.
That is, the groups of oscillators have sizes $n_1\neq n_2$. From these figures a number of features
are immediately obvious. The most striking feature is the ``plateau'' in the entanglement that is reached
whenever both groups are sizeable enough (say $n_1,n_2>10$, at least in the presented case for
coupling strength $\alpha=20$). Of course, when $n_1=n_2$, being the ``diagonal'' of the plot, we
recover the result of Section \ref{sec_symm} that the log-negativity is independent of $n=n_1+n_2$.
From these figures we are led to conjecture that, in the case of nearest-neighbor coupling,
the value of log-negativity for $n_1=n_2$ is an upper bound on
the values for $n_1\neq n_2$ (not to be confounded with the result of Theorem \ref{th_1}, which says that this
value is a {\em lower} bound for all symmetric bisections with general circulant couplings).
Moreover, for general circulant couplings, we conjecture that an upper bound on $N(n_1,n_2)$ is given by
$\lim_{m\rightarrow\infty} N(m,m).$
Another feature is that when, say, $n_1$ is kept fixed the log-negativity decreases with $n_2$ from
a given value of $n_2$ onwards. This phenomenon is seen most clearly with small $n_1$, particularly for
$n_1=1$. We will endeavour an intuitive explanation of these features below.
In conclusion, we conjecture that, again for general circulant couplings,
$\lim_{n_2\rightarrow\infty} N(1,n_2)$ is a lower bound on $N(n_1,n_2)$.

\begin{figure}
\leavevmode
\epsfxsize=8cm
\epsfbox{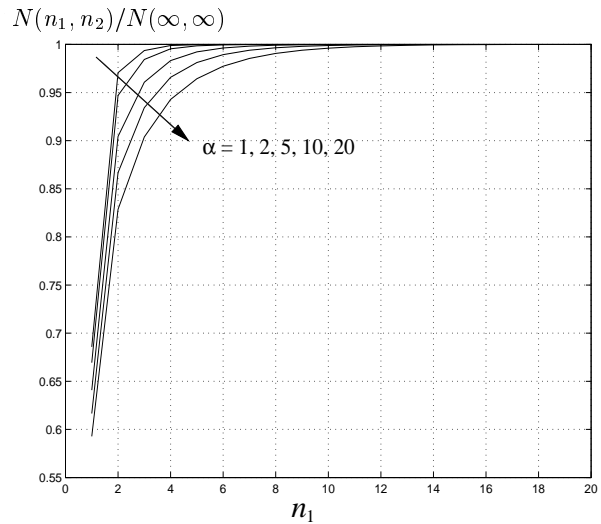}
\caption{Effect of coupling strength $\alpha$ on the convergence of the log-negativity towards its
maximal value. Group size $n_2$ is kept fixed at 20 and group size $n_1$ is varied. Shown is the ratio
$N(n_1,n_2)/N(\infty,\infty)$. The different curves are for various values of $\alpha$. One clearly sees
that for small couplings the limit value is reached much faster.}
\label{fig_plot8}
\end{figure}

From Figure \ref{fig_plot8} we can see that the convergence of $N$
towards its plateau value $N(\infty,\infty)$ depends on the
strength of the coupling $\alpha$. For higher values, convergence
is slower. What cannot be seen from this figure is that the actual
plateau value is larger as well.
\subsection{Entanglement versus energy}
It is interesting to compare the entanglement present in the chain
ground state with its energy. We consider nearest-neighbor
interaction only. We have shown in Section \ref{sec_cov} that the ground state
energy equals $(\hbar\omega/2)\trace [V^{1/2}]$. For zero coupling
($V=\identity_n$) this gives just $n$ times the single-oscillator
ground state energy $E_0=\hbar\omega/2$, as expected. For large
couplings $\alpha$, we show that the ground state energy is of
the order of $\sqrt{\alpha}n E_0$.

From the proof of Theorem \ref{th_2}, we have that the eigenvalues
$\lambda_k$ of $V$ are given by
\beas
\lambda_k = v_0+2v_1\cos(k
2\pi/n)+2v_2\cos(2k 2\pi/n)+\ldots.\eeas
The energy in terms of these
eigenvalues is $\sum_{k=0}^{n-1} \lambda_k^{1/2}$. For large
values of $n$, we can replace the discrete sum over $k$ by an
integral in $x=2\pi k/n$. For nearest-neighbor coupling, this
yields: \beas E &\approx& E_0 (n/\pi) \int_0^\pi dx (v_0+2v_1 \cos
2x)^{1/2} \\ &=& 2 E_0 (n/\pi) \int_0^{\pi/2} dx ((v_0+2v_1)-4v_1
\sin^2 x)^{1/2} \\ &=& 2 E_0 (n/\pi) \int_0^{\pi/2} dx (1+4\alpha
\sin^2 x)^{1/2}  \\ &=& 2 E_0 (n/\pi) \sqrt{\alpha} \int_0^{\pi/2}
dx (4\sin^2(x)+1/\alpha)^{1/2}.  \eeas

\begin{figure}
\leavevmode
\epsfxsize=8cm
\epsfbox{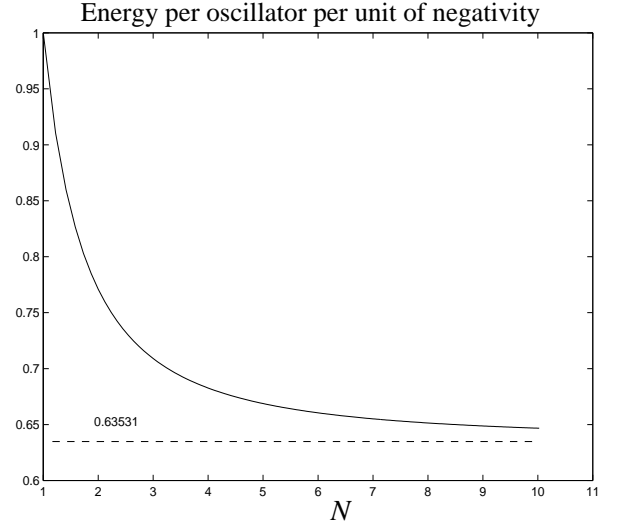}
\caption{Energy per oscillator (in units of $\hbar\omega/2$)
per unit of negativity (not logarithmic) in function of the
negativity, for the case of contiguous groups of large enough size (so that the entanglement
plateau in Figure \protect\ref{fig_plot2} is reached).
The interaction is nearest-neighbor and the coupling is implicitly present as a parameter.
The dashed line depicts the limiting value for infinitely strong coupling.
Here, the number $n$ of oscillators is taken to be 20.
However, the results become independent of $n$ for $n$ large enough:
for $n=20$, the limiting value is 0.63531, while for infinite $n$ the exact result is $2/\pi = 0.63662$.}
\label{fig_plot3}
\end{figure}

In the limit of $\alpha$ tending to infinity, the latter integral
tends to $2\int_0^{\pi/2} dx \sin(x)=2$, so that indeed \be E
\approx n E_0 \frac{4}{\pi} \sqrt{\alpha}. \ee Recalling the exact
formula for the log-negativity in the symmetrically bisected case,
we have that the negativity (not the logarithmic one) is
$\sqrt{1+4\alpha}$. We thus find that the negativity is
approximately proportional to the mean energy per oscillator. The
exact values, calculated numerically, have been plotted in Figure
\ref{fig_plot3}. For $\alpha=0$, the curve obviously goes through
the point with mean energy equal to $E_0$ and negativity equal to
1. For $\alpha$ going to infinity, the mean energy goes to
$(2/\pi)E_0 = 0.63662 E_0$ times the negativity.

\subsection{Non-contiguous groups}
\begin{figure}
\leavevmode
\epsfxsize=8cm
\epsfbox{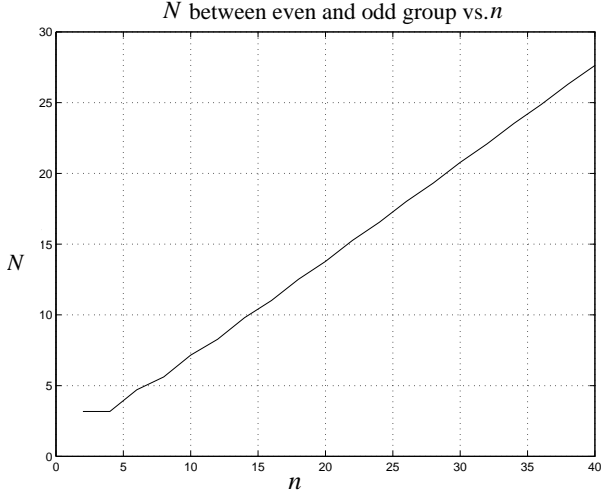}
\caption{Entanglement between the group of even oscillators and the group of odd oscillators, in function
of the chain length $n$ (even $n$ only). Interaction is again nearest-neighbor with coupling $\alpha=20$.
The log-negativity is seen to quickly converge to a constant times $n$. The value of the constant depends
on $\alpha$ and the relationship is shown in Figure \protect\ref{fig_plot6}.}
\label{fig_plot5}
\end{figure}

From the above, one would get the impression that the mean energy gives a general upper bound on the amount
of entanglement in the system (apart from a numerical factor). This is certainly not the case, because,
until now, we only have investigated the cases where the two groups of oscillators were contiguous.
In the following paragraph we look into the entanglement between non-contiguous groups. Specifically, we look
at the extreme case of entanglement between the
group of even oscillators and the group of odd ones.
As can be seen from
Figure \ref{fig_plot5}, numerical calculations already show that the log-negativity tends to a constant times $n$,
the chain length. Therefore, in this case, the log-negativity can grow indefinitely large even when the mean energy is
kept fixed. In view of this, it would be more correct to say that there are two contributions to the entanglement:
one is the mean energy, which is directly related to the coupling strengths, and the second is the surface
area of the boundary between the two groups of oscillators, which in the 1-dimensional case is just the number
of points where the two groups ``touch'' each other. We will return to this issue in Section \ref{sec_disc}.

The validity of the purported linear relationship can be shown analytically in a rather simple way,
yielding as a by-product an expression for the proportionality constant.
First of all, the diagonal elements of the $P$ matrix for this configuration are 1 for odd index values, and
-1 for even index values. The Fourier transform of $P$, that is: $\Omega P\Omega^\dagger$ (see proof of
Theorem \ref{th_2}),
is equal to \be\twomat{0}{\identity_{n/2}}{\identity_{n/2}}{0},\ee
as is easily checked.
We already have calculated the Fourier transform of the $V$ matrix,
which again can be inferred from the proof of Theorem \ref{th_2}. It is given by
$\Lambda=\Omega V \Omega^\dagger$; here,
$\Lambda$ is a diagonal matrix with diagonal elements
$\Lambda_k = v_0+2v_1\cos(2 k \pi/n)+2v_2\cos(4 k \pi/n)+\cdots$, $0\le k\le n-1$.
Inserting this in the expression for the $Q$-matrix gives:
\be
Q = \Omega^\dagger \Lambda^{-1/2} \twomat{0}{\identity_{n/2}}{\identity_{n/2}}{0}
\Lambda^{1/2} \twomat{0}{\identity_{n/2}}{\identity_{n/2}}{0} \Omega.
\ee
If we write $\Lambda$ in $2\times 2$ block form as
\be\twomat{\Lambda'}{0}{0}{\Lambda''},\ee
the spectrum of $Q$ is
the union of the spectrum of ${\Lambda'}^{-1/2}
{\Lambda''}^{1/2}$ and of ${\Lambda''}^{-1/2}{\Lambda'}^{1/2}$.
Worked-out, this gives the
eigenvalues $(\Lambda_{k+n/2}/\Lambda_k)^{1/2}$ and $(\Lambda_{k}/\Lambda_{k+n/2})^{1/2}$,
for $0\le k\le n/2-1$. Using the inherent symmetry that $\Lambda_{n-k}=\Lambda_k$, the eigenvalues of $Q$ are
\be
\left(\frac{\Lambda_k}{\Lambda_{n/2-k}}\right)^{\pm 1/2}.
\ee
The formula for the log-negativity obtained in Section \ref{sec_logneg}
can be reformulated as minus the sum of the negative
eigenvalues of $\log_2 Q$. In the present case we get as log-negativity
\be
N = \frac{1}{2}\sum_{k=0}^{n/2} |\log_2 \frac{\Lambda_k}{\Lambda_{n/2-k}}|.
\ee

\begin{figure}
\leavevmode
\epsfxsize=8cm
\epsfbox{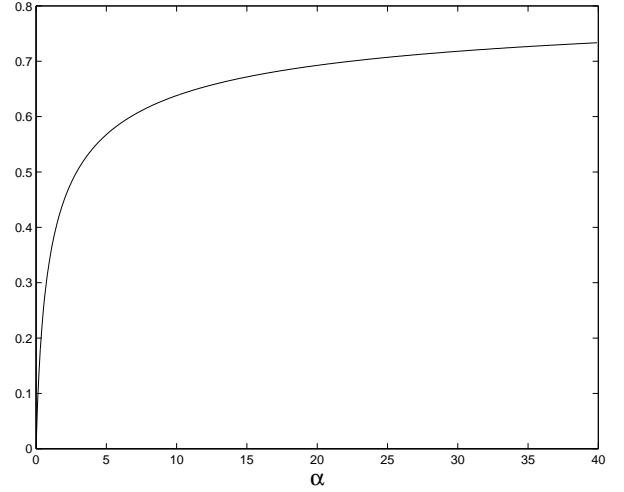}
\caption{Relationship between the constant factor $c$ appearing in the asymptotic formula for the log-negativity
$N=cn$ in the even-odd setting (as in Figure \protect\ref{fig_plot5}) and the coupling constant $\alpha$.}
\label{fig_plot6}
\end{figure}

For the nearest-neighbor Hamiltonian, this simplifies to
\beas
N &=& \frac{1}{2}\sum_{k=0}^{n/2} |\log_2 \frac{v_0+2v_1\cos(2\pi k/n)}{v_0-2v_1\cos(2\pi k/n)}| \\
&=& \sum_{k=0}^{n/4}
\log_2 \frac{1+2\alpha(1+\cos(2\pi k/n))}{1+2\alpha(1-\cos(2\pi k/n))}
\eeas
for $n$ that are multiples of 4.
For large $n$, we can replace the discrete sum by an integral,
\be
N \approx \frac{n}{2\pi} \int_0^{\pi/2} dx \log_2 \frac{1+2\alpha(1+\cos(x))}{1+2\alpha(1-\cos(x))},
\ee
which indeed proves that, for large $n$, the log-negativity is a linear function of $n$.
The integral itself cannot be brought in closed form. In Figure \ref{fig_plot6},
we show the result of numerical calculations giving the asymptotic value of $N/n$ versus $\alpha$.

\subsection{Effect of group separation}
\begin{figure}
\leavevmode
\epsfxsize=8cm
\epsfbox{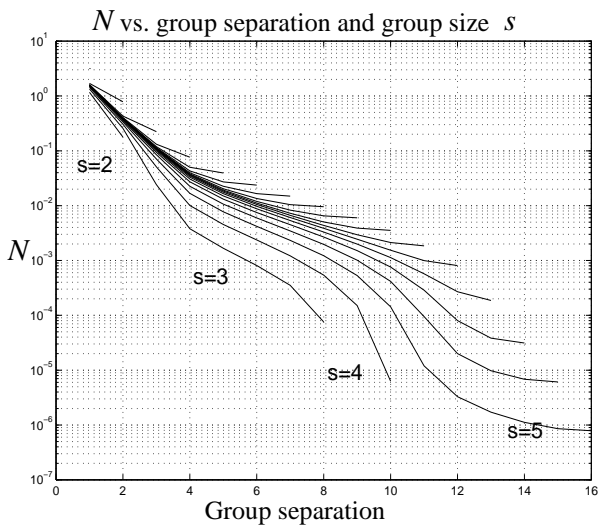}
\caption{Log-negativity for two contiguous groups that do not comprise the whole chain.
The chain consists of 40 oscillators, coupling is nearest-neighbor with coupling strength
$\alpha=20$. Shown is the log-negativity, displayed on a logarithmic scale, versus the separation between
the groups, i.e.,  the number of oscillator positions between them.
The different curves are for different group sizes $s$ (both groups are taken to be equal in size).
The curve for group size 1 is not visible because it is a single point:
the log-negativity between 2 oscillators turns out to be 0
whenever their separation is larger than 0.}
\label{fig_plot7}
\end{figure}

In the following paragraph, we give some results for contiguous groups that do not comprise the whole chain.
In Figure \ref{fig_plot7} we consider a fixed chain of $n=40$ oscillators and look at the entanglement between
two equally sized contiguous groups, in function of the group size and the separation between them. We define
the separation as the number of oscillators in the smallest gap between the groups; since we are dealing with a
ring, there are two gaps between the groups. Note that the log-negativity is plotted on a logarithmic scale.
There are two main features in this figure. The first and least unexpected feature is that the entanglement
decreases more or less exponentially with the separation. We believe that this is quite natural in view
of the fact that the coupling between the groups also decreases with the distance.

The more remarkable feature is that for small groups, the
entanglement quickly becomes zero altogether, as measured
by the logarithmic negativity. Bound entanglement \cite{Bound,Boundcv}
is of course not detected by this measure of entanglement,
and it would be an interesting enterprise in its own
right to study the structure of bound entanglement present
in coupled oscillator systems. We will leave this, however,
for future investigations. From now on, the term that
no entanglement is present will be
used synonymically with the statement
that the logarithmic negativity vanishes.

For groups of size 1, the log-negativity
is zero already at separation 1 (the
gap consists of one oscillator). For groups of size 2, there is still entanglement at separation 1, but none
at separation 2. The larger the groups, the larger the maximal separation for which there is still entanglement
can be. One could try to interpret this by saying that there is a kind of threshold value below which
entanglement drops to zero. However, this is more a reformulation of the results than an explanation, because
it sheds no light on why this supposed threshold should depend on the group size.

To really explain what is happening, we need to take a closer look at the exact calculations.
Consider first two groups of oscillators of size 2 and with separation 1; that is, group 1 is at positions
1 and 2, group 2 at 4 and 5. The $Q$-matrix
of this configuration (with $n=40$ and $\alpha=20$) has eigenvalues
        2.063,
       1.1339,
       1.0938 and
      0.88361.
As one of the eigenvalues is smaller than 1, there is, indeed, entanglement
present.
We might be led to think that this entanglement is the cumulative result of the entanglement
between the different oscillator pairs, (1,4), (1,5), (2,4) and (2,5), but this is not true,
because these pairs are not entangled themselves: their separation is larger than 0.
What is happening here is that the eigenvalues of the $Q$-matrix belonging to pair (1,4), say,
are 1.8065 and 1.1724, which are both larger than 1 and, therefore, do not count in the entanglement
figure.

The resolution of this strange behaviour in terms of the separation is that the mere fact alone of
having correlations between the groups (eigenvalues of $Q$ different from 1) is not enough to have
entanglement. The correlations must be of special nature, namely: the eigenvalues of $Q$ must be smaller than 1.
One could say that larger groups can more easily exhibit entanglement; their $Q$ matrix has a larger
dimension and, hence, more eigenvalues, so that there are more opportunities for having at least one eigenvalue
smaller than 1.

\begin{figure}
\leavevmode
\epsfxsize=8cm
\epsfbox{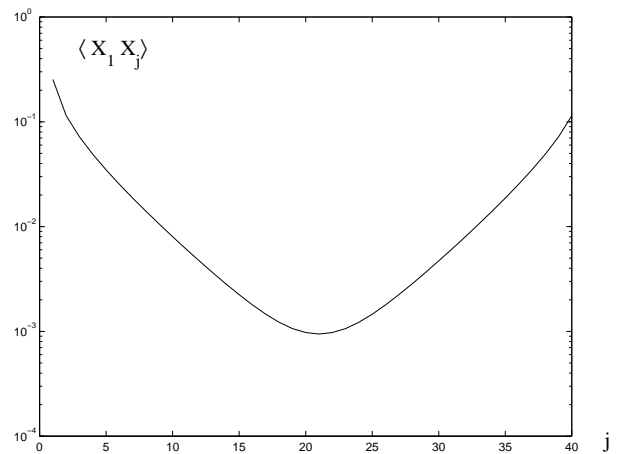}
\caption{Classical correlations in a chain consisting of 40 oscillators;
coupling is nearest-neighbor with coupling strength
$\alpha=20$. Shown is the quantity $\langle \hat X_1 \hat X_j\rangle$, displayed on a logarithmic scale,
versus the second oscillator index $j$.}
\label{fig_plot10}
\end{figure}

In this respect, it is interesting also to have a look at the {\em
classical} correlations in the chain, i.e.\ the expectation values
$\langle \hat X_j \hat X_k\rangle$. From the treatment in Section \ref{sec_cov}
we immediately see that these correlations are given by the
elements of the matrix $\gamma_x/2 = V^{-1/2}/2$. In the case of
circulant symmetry, we only have to consider the first row of the
matrix, giving the correlations between the first oscillator and
any other one. Figure \ref{fig_plot10} shows these classical
correlations for the system considered in Figure \ref{fig_plot7}
($n=40$, $\alpha=20$). As could be expected, these correlations
decrease exponentially with the oscillator distance and,
furthermore, never vanish completely.
\subsection{Thermal state}
\begin{figure}
\leavevmode \epsfxsize=8cm \epsfbox{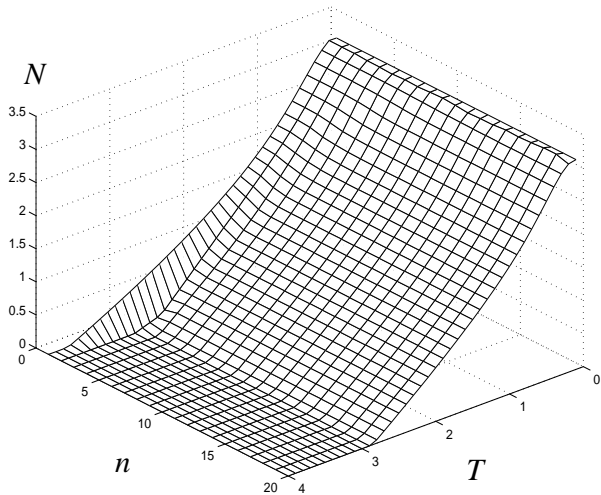}
\caption{Log-negativity of a thermal state with temperature $T$
versus $T$ and chain size $n$. Symmetrically bisected chain,
nearest-neighbor interaction with coupling $\alpha=20$.}
\label{fig_plotth}
\end{figure}

To conclude this section, we consider a thermal state instead of the ground state.
The calculations are exactly the same in both cases, apart from the fact that in the covariance matrix
there is an additional factor $\identity_{n} + 2(\exp(\beta \sqrt{V})-\identity_{n})^{-1}$ to the $\gamma_p$ and
$\gamma_x$ blocks ($T=1/\beta$).
The results are shown in Figure \ref{fig_plotth}.
One sees that for small temperatures the negativity is equal to the ground state negativity, and from some
value onwards it starts to decrease more or less linearly with $T$ until there is no (free)
entanglement at all anymore.
\section{Discussion}\label{sec_disc}
The numerical results we have obtained in Section \ref{sec_gen} can be interpreted in a qualitative way,
by means of two rules-of-thumb. These rules are not to be interpreted as strict mathematical statements;
for that, we already have the exact formulas. The importance of the two rules is that they allow
to reason about the dependence of entanglement on various factors, like group size, coupling strengths
and group geometry.

\begin{figure}
\leavevmode
\epsfxsize=6cm
\centerline{ \epsfbox{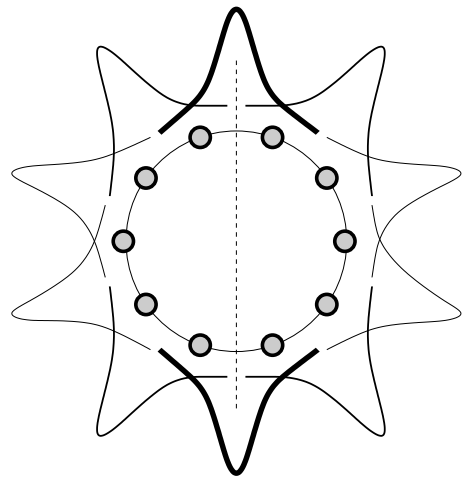}}
\caption{Schematic drawing of the inter-oscillator correlation
packets, i.e.\ the rows of the correlation matrix $\gamma_x$. The line thickness indicates the amount by which
the correlation packet is involved in the entanglement between the
groups, i.e., how much it is shared by the two groups.}
\label{fig_plotrob}
\end{figure}

The first rule is that, {\em due to the coupling between the oscillators, the system exhibits inter-oscillator
correlations which are decreasing with distance}.
This is a fairly natural statement, in view of the fact that the couplings between the oscillators
are short-range as well. In a more mathematical way, one could
consider the matrix $\gamma_x = V^{-1/2}$, whose elements are the classical
correlations $\langle \hat X_j \hat X_k\rangle$.
The $j$-th row describes the correlations
between the $j$-th oscillator and all other oscillators.
The correlations can thus, in a figurative way, be subdivided
into packets, one packet for every row in the correlation matrix.
For chains with a circulant potential matrix $V$, it is self-evident that $\gamma_x$
is also circulant so that the correlation packets all have an identical shape
(see Figure \ref{fig_plotrob}).

The second rule is that {\em the entanglement between two
groups of oscillators depends on the total amount of correlation between
the groups}. Again, this rule looks fairly innocuous and even trivial. However,
combining the two rules readily shows why, in the case of contiguous groups, the entanglement in function of
the group sizes should reach a plateau.
Indeed, even while the total amount of correlation grows, more or less, linearly with the chain size,
this has very little impact on the entanglement between the groups because it are only the
correlation packets that straddle the group boundaries that enter in the bipartite entanglement figure.
For large groups, most of the correlation packets describe correlations {\em within} the groups.
What is important is the amount of correlations {\em between} the groups, and this quantity is virtually independent
on the group size, provided the groups are so large they can accomodate most of the packets within
their boundaries.

We must stress, however, that these two rules are of a qualitative nature.
As noted already in Section \ref{sec_gen},
in the discussion of the dependence of entanglement on group separation,
having correlations between the
groups alone is not enough for having entanglement. The correlations must be
such that the $Q$ matrix has at least
one eigenvalue smaller than 1.
The bottom line is in any case
that one must go through the exact calculations to see whether or not there
is entanglement.

Another effect that can be accounted for is the dependence of the log-negativity on the group size if at least
one group is very small. If both groups are
very small, say 1 oscillator both, then the packets are so wide they wind up along the chain
and, therefore, cross every group more than once, adding to the entanglement figure a number of times.
If one of the groups is kept fixed, and the other
is made larger, the winding number of the packets decreases and so does the amount by which the packet
is shared by the groups. This effect could explain the decrease of entanglement with growing $n_2$.
At this point, the qualitative reasoning again breaks down, however, since the reduction of the
amount of sharing per packet is counteracted by an increase in the number of packets. To show that the
balance is still in favour of an entanglement decrease, once again one really needs to go through the exact
calculations; this is what we have done in Section \ref{sec_symm}.
Nevertheless, the qualitative reasoning has the virtue that it shows what the main ingredients are.
Furthermore, it immediately leads to the conjecture that the effect of decreasing entanglement
would not occur in a chain that is not connected end-to-end, since no winding occurs there.

\begin{figure}
\leavevmode
\epsfxsize=8cm
\epsfbox{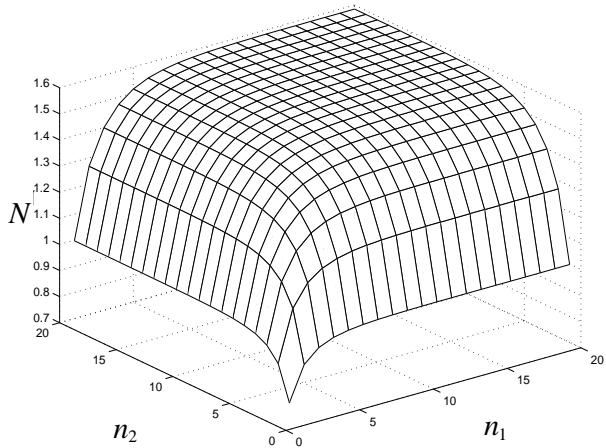}
\caption{Same as
Figure \protect\ref{fig_plot2}, but with a Hamiltonian that does
not connect the harmonic chain end-to-end. The two groups are,
therefore, connected only at one point (in the middle of the
chain), and this explains why the log-negativity is only about one
half of the value it had with end-to-end connection. Furthermore,
for small $n_1$ we now see an increase with $n_2$ instead of a
decrease, which seems to imply that the counterintuitive behaviour
on the ring is actually a winding effect (see text).}
\label{fig_plot4}
\end{figure}

A numerical experiment immediately showed that this is exactly what happens,
as witnessed by Figure \ref{fig_plot4}.
One has to be careful, though, about how one ``opens'' the chain.
To clearly show the disappearance of the winding
effect, one has to make sure that opening the chain does not
introduce side-effects. Particularly, the oscillators at both ends
should still ``see'' the same springs as before opening the chain.
One can take care of this by connecting the ends of the chain to
two additional oscillators that are kept in a zero-energy state (i.e.\ with
zero $X$-variance; hence, they must be oscillators with infinite mass).
At the level of the potential matrix, this
means that the diagonal elements $V_{1,1}$ and $V_{n,n}$ are still
$1+2\alpha$, although the elements $V_{1,n}$ and $V_{n,1}$ are being set to
zero. Noting the analogy between harmonic chains and
transmission lines, we call this special connection process the
{\em termination} of the harmonic chain. In transmission line
theory, correct termination of a line (using appropriately matched impedances)
is necessary to avoid signal
reflections at the ends of the line. We believe that analogous reflection
effects could be exhibited by non-terminated harmonic chains,
but leave the investigation of this boundary phenomenon to future work.

Finally, for non-contiguous groups, and, specifically, for the
entanglement between even and odd oscillators, the two
rules-of-thumb correctly predict that the entanglement keeps
increasing with growing chain length. Indeed, if $n$ grows, then
the ``boundary area'' between the even and odd group also grows
(linearly with $n$), in contrast with the contiguous groups, whose
boundary area is fixed (1 for the open chain, 2 for the closed
chain). Hence, the amount of correlations straddling the boundary
should grow too. The exact calculation confirms this effect and
shows a linear relationship between log-negativity and chain size.
It would be interesting to investigate what happens in
three-dimensional oscillator arrangements with couplings
decreasing with distance. We believe that a similar relation will
show up between entanglement between two groups and the area of
the boundary between the groups. We leave this issue, however, for
future investigations.
\begin{acknowledgments}
This paper was benefited from very interesting conversations with
C. Simon, R. Ratonandez, S. Scheel, M. Santos, and J. Harley. 
This work was supported by
the European Union project EQUIP, the European Science Foundation
programme on ``Quantum Information Theory and Quantum Computing'', by
a grant of the UK Engineering and Physical Sciences Research Council
(EPSRC), and by a Feodor-Lynen grant of the Alexander von Humboldt
Foundation.
\end{acknowledgments}

\end{document}